\documentclass[smallextended, 11pt, twocolumn]{IEEEtran}  
\usepackage[numbers,sort&compress]{natbib}
\usepackage{tikz}
\usepackage{hyperref}
\usepackage{graphicx}  
\usepackage{qcircuit}

\usepackage{caption}
\usepackage{subcaption}
\usepackage{amsmath,amsfonts,amsthm, bm}

\everymath{\displaystyle} 
\usepackage{colortbl}

\usepackage{todonotes}
\newcommand{\ket}[1]{\ensuremath{\left|#1\right\rangle}} 

\renewcommand{\bf}[1]{\ensuremath{\mathbf{#1}}}
\begin{document}

\title{{ Quantum implementation of circulant matrices and its use in quantum string processing}


\author{Ammar Daskin*
\IEEEauthorblockN{
              \thanks{*adaskin25@gmail.com}           
}
}
\date{Received: date / Accepted: date}
}


\maketitle

\begin{abstract}

String problems in general can be solved faster by using special data structures such as suffixes in many cases structured as trees and arrays. In this paper, we show that suffixes used in those data structures can be obtained by using circulant matrices as a quantum operator which can be implemented in logarithmic time. Hence, if the strings are given as quantum states, using the presented circuit implementation one can do string processing efficiently on quantum computers.
\end{abstract}

\section{Introduction}

For any given vector $\bf{c} = [c_0, c_1, \dots, c_{n-1}]^T$, the circulant matrix $C$ with the first row $\bf{c}$ is defined as follows (\cite{golub2013matrix,gray2006toeplitz,karner2003spectral}):
\begin{equation}
\label{Eq:C}
C = \left(\begin{matrix}
c_0      & c_{n-1} & \cdots  & c_2     & c_1     \\
c_1      & c_0     & c_{n-1} &         & c_2     \\
\vdots   & c_1     & c_0     & \ddots  & \vdots  \\
c_{n-2}  &         & \ddots  & \ddots  & c_{n-1} \\
c_{n-1}  & c_{n-2} & \cdots  & c_1     & c_0     \\
\end{matrix}\right)
\end{equation}
The eigenspace of the circulant matrices are formed by Fourier matrices; therefore, the eigenvectors and the eigenvalues of $C$ are defined analytically in the following form of pairs:
A $j$th ($j = 0, \ldots, n-1$) eigenvector is given by
\begin{equation}
v_j=\frac{1}{\sqrt{n}} \left(1, \omega^j, \omega^{2j}, \ldots, \omega^{(n-1)j}\right),
\end{equation}
where $\omega=\exp \left(\tfrac{2\pi i}{n}\right)$.
And its corresponding eigenvalue is
\begin{equation}
\label{Eq:EigenC}
\lambda_j = c_0+c_{n-1} \omega^j + c_{n-2} \omega^{2j} + \dots + c_{1} \omega^{(n-1)j}.
\end{equation}
As a result, the eigendecomposition of any circulant matrix can be written as $C = F\Lambda F^{-1}$ where $F$ represents the normalized matrix for the discrete Fourier transform and $\Lambda = diag({F \bf c})$, which is a diagonal matrix representing the eigenvalues. 

\subsection{Direct Implementation as a quantum circuit}

One can implement $C$ as a quantum circuit by using its eigendecomposition $C = F\Lambda F^{-1}$: It is well known that $F$ can be implemented in $O(\log(n)^2)$ time as the quantum Fourier transform \cite{nielsen2002quantum}. 
In addition, since any vector of dimension $n$ can be implemented in $O(n)$ time, we can also implement $\Lambda$ in linear time. Therefore, $C$ can be implemented as a quantum circuit by using $O(n)$ two and single qubit quantum gates, which is less than the classical matrix size $O(n^2)$. 

\subsection{Circuit implementation through permutations}
$C$ can be represented as a function of the following permutation matrix, which is also circulant:
\begin{equation}
\label{EqMatrixP}
P = \left( \begin{matrix}
 0&0&\cdots&0&1\\
 1&0&\cdots&0&0\\
 0&\ddots&&\vdots&\vdots\\
 \vdots&\ddots&\ddots&0&0\\
 0&\cdots&0&1&0
\end{matrix} \right)
\end{equation}
Then, we can redefine $C$ as a polynomial function of $P$ with the coefficients given by $\bf{c}$: 
\begin{equation}
C = c_0 I + c_1 P + c_2 P^2 + \dots + c_{n-1} P^{n-1} = \sum_{j=0}^{n-1}c_jP^j.
\end{equation}
Since $P$ is an orthogonal matrix, we can construct its exact quantum circuit decomposition. 
Moreover, it is known that sum of unitary matrices can be implemented as a subpart of a larger quantum  system with the help of ancilla registers (e.g.,  \cite{childs2012hamiltonian,daskin2012universal,low2019hamiltonian}).  
Here, we shall show that $C$ can be constructed as a vector by using $V_P\times \ket{\psi}$ where \ket{\psi} represents a combinations of the vector $\bf{c}$ and $V_P$ includes the powers of $P$ used in the above polynomial. 
Later in the paper, we shall show that this way of implementation eases the way to solve some string problems. 

\subsubsection{Circuit for $P$}
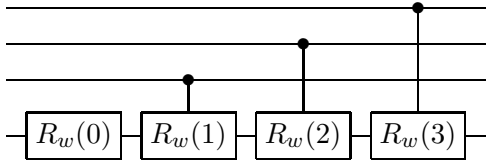
\begin{figure}[t]
	\begin{center}
\colorbox{white!1}{
		\Qcircuit @C=0.7em @R=1em {
\lstick{} &\qw &\qw &  \qw &\ctrl{3} & \qw & \\
\lstick{} &\qw &\qw& \ctrl{2}  & \qw &  \qw &   \\
\lstick{} & \qw &\ctrl{1}&\qw & \qw &  \qw &   \\
\lstick{}&\gate{R_w(0)} &\gate{R_w(1)}  &\gate{R_w(2)}  & \gate{R_w(3)}&\qw   \\
			}
}
	\end{center}
	\caption{\label{FigCircuitLambdaP} Quantum circuit for $\Lambda_P$}
\end{figure}

Since $P$ is a circulant matrix, we can define its any $j$th eigenvalue by using Eq.\eqref{Eq:EigenC}:
\begin{equation}
\lambda_j = w^j.
\end{equation}
Therefore, its eigendecomposition can be written as:
\begin{equation}
\label{EqLambdaP}
F\times 
\left(\begin{matrix}
1&&&&\\
&w^1&&\\
&&w^2&\\
&&&\ddots&\\
&&&&w^{n-1}
\end{matrix}\right)\times F^{-1} =F \Lambda_P F^{-1} .
\end{equation}
As a quantum circuit, we can implement $\Lambda_P$ as follows:
 First, observe that the power $j$ in $w^j$ is equal to the row index which can be found from the binary expansion of $j = (b_{n-1}\dots b_0)_2$. Therefore, if a qubit is in \ket{1}, we apply a control phase gate to the first qubit. The phase of these gates are determined from the decimal index of the qubit in the binary expansion.
We define the phase gate as follows:
\begin{equation}
\label{Eq:Rw}
R_w(k) = \left(\begin{matrix}
  1 & 0\\
  0 & w^{2^k}
\end{matrix}\right),
\end{equation}
where $k$ is the order of the control qubit in the binary expansion $(b_{n-1}\dots b_0)_2$.
The resulting circuit is drawn in Fig.\ref{FigCircuitLambdaP} for four qubits. The number of quantum gates for this implementation is equivalent to the number of qubits which is $(\log n)$. 

\subsubsection{Circuit for $V_P$}
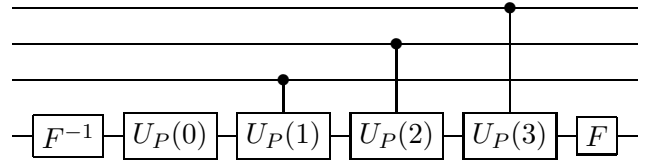
\begin{figure}[t]
	\begin{center}
{
		\Qcircuit @C=0.7em @R=1em {
\lstick{} &\qw&\qw &\qw &  \qw &\ctrl{3} & \qw & \qw\\
\lstick{} &\qw&\qw &\qw& \ctrl{2}  & \qw &  \qw & \qw  \\
\lstick{} &\qw& \qw &\ctrl{1}&\qw & \qw &  \qw & \qw  \\
\lstick{}&\gate{F^{-1}}&\gate{U_P(0)} &\gate{U_P(1)}  &\gate{U_P(2)}  & \gate{U_P(3)}&\gate{F}&\qw   \\
}}
	\end{center}
	\caption{\label{FigCircuitV} The final quantum circuit of $V_P$ illustrated for $n=4$. The explicit circuit for $U_P(k)$ is given in Fig.\ref{FigUpk} below.}
\end{figure}

\begin{figure}[t]
{
		\Qcircuit @C=0.7em @R=1em {
\lstick{} &\ctrl{4}&\ctrl{4} &\ctrl{4} &  \ctrl{4}  & \qw & \\
\lstick{} &\qw &\qw &  \qw &\ctrl{3} & \qw & \\
\lstick{} &\qw &\qw& \ctrl{2}  & \qw &  \qw &   \\
\lstick{} & \qw &\ctrl{1}&\qw & \qw &  \qw &   \\
\lstick{}&\gate{R_w(0\times k)} &\gate{R_w(1\times k)}  &\gate{R_w(2\times k)}  & \gate{R_w(3\times k)}&\qw    \\
}}
	\caption{\label{FigUpk} The explicit circuit implementation of $U_P(k)$  which implements the $2^k$th power of $\Lambda_P$ given in Eq.\eqref{EqLambdaP}.}
\end{figure}
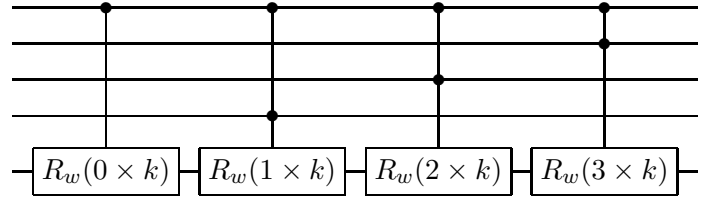

Consider the following matrix:
\begin{equation}
V_P = 
\left(\begin{matrix}
I&&&&\\
&P^1&&\\
&&P^2&\\
&&&\ddots&\\
&&&&P^{n-1}
\end{matrix}\right),
\end{equation}
which highly resembles $\Lambda_P$ given in Eq.\eqref{EqLambdaP}. Therefore, we can implement the same way as we implemented $\Lambda_P$: In this case,  we only need to somehow replace $R_w({k})$ with the power of $P$: 
\begin{equation}
P^{2^k}=F\Lambda_P^{2^k}F^{-1}.
\end{equation} 
Since all the powers of $P$s have the same eigenvectors, we apply $F$ and $F^{-1}$ only once at the beginning and at the end of the circuit. We use the following quantum operation in the circuit:
\begin{equation}
U_P(k) = \left(\begin{matrix}
  I & 0\\
  0 & \Lambda_P^{2^k}
\end{matrix}\right).
\end{equation}
Here,  when we look at the circuit for $\Lambda_P$ in Fig.\ref{FigCircuitLambdaP}, we can see that the power can be distributed to the quantum gates inside the circuit since the order of the gates is not important. 
That means, we can obtain any power of $\Lambda_P$ by simply adjusting the angle values of the quantum gates. 
Therefore, for power of $\lambda_P^{2^k}$, we adjust the parameter of the rotation gates in Fig.\ref{FigCircuitLambdaP} as: $R_w(0\times k), R_w(1\times k), \dots$, and so on. The final circuit for $V_P$ is just composed of the controlled versions of these quantum gates which is shown in Fig.\ref{FigCircuitV} and \ref{FigUpk}.

\subsection{The overall complexity}
As shown in Fig.\ref{FigCircuitV} and \ref{FigUpk} the number of quantum gates are limited to the number of qubits used in the main system. The size of the permutation matrix $P$ is $n$ by $n$, therefore the main system is described by $\log n$ qubits. Then, an additional $\log n$ qubit is used to construct $V_P$. Therefore, in total the circuit requires $(2\times \log n)$ qubits.
The quantum gates are at most controlled by two qubits, therefore the total number of required single and two qubit gates are bounded by $O(poly(\log n))$ in general. More specifically, if we assume the general implementation of quantum Fourier transform requires $O((\log n)^2)$ gates (with an optimization, it may require $O((\log n)\log\log n)$ ) and the decomposition of each Toffoli gate requires less than $\log n$ gates \cite{nielsen2002quantum}, then the total complexity becomes bounded by $O((\log n)^2)$.
This shows that we can form $V_P$ on quantum computers very efficiently since it requires   a number of quantum gates which is a poly-logarithm of the dimension $n$.

\section{Applications}
\subsection{Suffix trees and arrays}
Many string problems can be solved easier if they are stored by using well-known data structures such as suffix tries, trees, and arrays \cite{manber1993suffix}. Below, we will follow the lecture notes in Ref.\cite{langmeadburrows} to give a brief introduction of these data structures and explain how to implement them on quantum computers with the help of the circuits introduced in the previous section.

From a given string, a suffix can be chosen by determining a starting position. For instance, we can choose the following suffixes from string \textbf{``banana"}:
\begin{equation}
\begin{bmatrix}
b&a&n&a&n&a&\$\\
a&n&a&n&a&\$\\
n&a&n&a&\$\\
a&n&a&\$\\
n&a&\$\\
a&\$\\
\$
    \end{bmatrix}
\end{equation}
Here, ``\$" is used to indicate the end of the string and   considered less than any letter of the alphabet in lexicographical order.

In suffix arrays, these strings are put in the buckets based on their orders and sorted. 
In tries, the prefixes are put in the rooted tree in which every node has a child for each letter of the alphabet used in the string. 
In the suffix tries, the suffixes are put in a rooted tree in which every node has a child for each letter of the alphabet used in the string. 
Therefore, by following the paths from the root of the tree, it is possible to determine if a letter follows another and if the prefix exists or not.
In the suffix trees, the tries are built for the suffixes and the number of nodes in the tries are optimized by compressing the nodes in the straight paths ( i.e. the paths that do not go more than one way in any nodes on the path.). 
Each path in the suffix tree ends with the letter ``\$".

While constructing suffix arrays and trees  can be done in $O(n\log n)$ time \cite{ukkonen1995line}, they can be used to solve many string problems more efficiently such as string searching and matching and sequence alignment \cite{mielczarek2016review}. 

\subsection{Burrows Wheeler Transform\cite{burrows1994block}}

The main idea in Burrows Wheeler transform (BWT) is to first  generate a circulant matrix by using the given string and then sort this matrix column by column. In the sorted matrix, the last column is used as the equivalent transformed string. As an illustration, consider the following example string "banana", which is used in most textbooks: 
\begin{equation}
    \begin{bmatrix}
b&a&n&a&n&a&\$\\
a&n&a&n&a&\$&b\\
n&a&n&a&\$&b&a\\
a&n&a&\$&b&a&n\\
n&a&\$&b&a&n&a\\
a&\$&b&a&n&a&n\\
\$&b&a&n&a&n&a
    \end{bmatrix} \xrightarrow{sort}
        \begin{bmatrix}
\$&b&a&n&a&n&\bf a\\
a&\$&b&a&n&a&\bf n\\
a&n&a&\$&b&a&\bf n\\
a&n&a&n&a&\$&\bf b\\
b&a&n&a&n&a&\bf \$\\
n&a&n&a&\$&b&\bf a\\
n&a&\$&b&a&n&\bf a\\
    \end{bmatrix} 
\end{equation}
From the above, BWT(``banana") = ``annb\$aa". In the above matrix, if one considers ``\$" as 1s and the rest of the letters as 0s, then BWT of any string can be represented as a permutation matrix. 
\subsection{Quantum implementation of suffixes}
In the previous section we show how to construct $V_P$. 
Here, consider that we are given the text \ket{c} including also the ``\$" sign.
We apply \ket{c} in the main register and $H^{\otimes \log n}\ket{0}$ in the ancilla register. This generates the following vector:
\begin{equation}
(H^{\otimes \log n}\ket{0}) \ket{c} = 
\frac{1}{\sqrt{n}}\left(\begin{matrix}
\begin{bmatrix}
 c_0\\c_1\\ \vdots \\c_{n-1}\\
\end{bmatrix}_{0\ \ \ }\\
\vdots\ \ \ \ \\
\begin{bmatrix}
 c_0\\c_1\\ \vdots \\c_{n-1}\\
\end{bmatrix}_{n-1}\\
\end{matrix}\right).
\end{equation}
If we multiply this vector by $V_P$, we get $C$ in Eq.\eqref{Eq:C} in vector form:
\begin{equation}
V_P (H^{\otimes \log n}\ket{0}) \ket{c} = 
\frac{1}{\sqrt{n}}\left(\begin{matrix}
\begin{bmatrix}
 c_0\\c_1\\ \vdots \\c_{n-1}\\
\end{bmatrix}_{0\ \ \ }\\
\vdots\ \ \ \ \\
\begin{bmatrix}
 c_{n-1}\\c_{n-2}\\ \vdots \\c_{0}\\
\end{bmatrix}_{n-1}\\
\end{matrix}\right) 
=  
\frac{1}{\sqrt{n}}\left(\begin{matrix}
S_{0}\\
\vdots \\
S_{n-1}\\
\end{matrix}\right),
\end{equation}
where each $S_j$ represents a circularly permuted string.
The above vector provides the unsorted suffixes whose end indicated by some end of file character "\$". 
Since the order of this character provides  information about the suffixes, the amplitude of this char may be adjusted higher so that from the measurements, the order can be obtained in an efficient way.

By using the indices of the end of file characters, we can collapse the quantum state onto any desired direction. 
In addition, we can draw some conclusion from the measurements on the collapsed state: e.g., . Therefore, the measurement statistics can be used to draw conclusions about the most common prefixes in the suffixes  in the collapsed state. 

Since in most string algorithms sorted structures allow us to develop more efficient algorithms, the same sorting can be also done on this vector.

\section{Sorting}
Sorting problem simply can be described as finding an ordered list of items given as an unordered list.
It is known that comparison based sorting algorithms such as merge sort and quicksort have running time $\Omega(n log n)$ for a list of $n$ items. 
This lower bound can be broken by using  bucket sorting (counting sort) where each item is considered as a direct pointer to the bucket; therefore, the sorting is done in $O(n)$ time. This simple approach is further improved to give $O(nd)$ time, where $d$ is the number of bits used to represent each item and $O(nlog log n)$ time. 
Because of the memory constraints and requirements,  in practice comparison sorts are used  more often in practice than these kinds of sorting algorithms. \cite{hagerup1998sorting}

On quantum computers, the sorting problem is considered based on registers $\ket{x_1} \dots \ket{x_n}$. Here, each register represents an item in the unsorted list. 
Therefore, the algorithms try to prepare the registers in the output to encode the natural ordered list of the given items: i.e., $\ket{q_1} \geq \dots \geq \ket{q_n}$.
Based on this representation, it is shown that quantum algorithms based on comparison models have similar complexity bounds for the sorting problem: i.e. $\Omega (\sqrt{n}logn)$ \cite{hoyer2002quantum,ambainis2002quantum}.

As done in the classical sorting algorithms, bucket sort with an order preserving hash function can be used to sort the items stored in memory as:
\begin{equation}
\sum_{i=1}^n\ket{\bf 0}\ket{i}\ket{x_i},
\end{equation}
where $i$ is the index of the item $x_i$ in the given unordered list.
The sorting problem becomes constructing the following quantum state:
\begin{equation}
\sum_{i=1}^n\ket{o_i}\ket{i}\ket{x_i}
\end{equation}
where \ket{o_i} represents the index of the item in the sorted list.
We can rewrite this in terms of a hash function $h(x_i)$ that maps an item $x_i$ to the index $o_i$:
\begin{equation}
\sum_{i=1}^n\ket{h(x_i)}\ket{i}\ket{x_i}.
\end{equation}
$h$ can be as simple as a direct map or more general hash function. 
In particular consider
$h$ as a partial order preserving hash function: i.e., if $x_i$ and $x_j$, their real ordered locations at some distance $d$ from each other so that $o_i + d < o_j$, then $h(x_i) < h(x_j)$.  
Then, since we can apply an operator $h$ simultaneously to all items, we can generate their orders in $O(1)$ time.
Sometimes knowing the elements' rough order in the array may be considered enough. 
In those cases,  the hash function need not be perfect; therefore,
one can use similar ideas to classical sorting algorithms such as bucket sort or the shell-sort \footnote{Shell sort is a generalization of insertion sort algorithm, where items at certain distances are compared and if necessary swapped to have a k-sorted array: i.e. an array where the numbers are grouped into regions based on their orders. } to generate some k-sorted array in which buckets are sorted, however, the items in the same buckets are not sorted. 

As mentioned above, if we use comparison based sorting algorithms, then the sorting is almost the same as classical sorting algorithm:  
Let us consider the following vector whose construction is given in the previous section:
\begin{equation}
    \frac{1}{\sqrt{n}}\left(\begin{matrix}
S_{0}\\
\vdots \\
S_{n-1}\\
\end{matrix}\right),
\end{equation}
In Burrows Wheeler transform, the items are sorted by columns.
We can do the same sorting on this vector: A particular direct sorting may be as follows: 

\begin{itemize}
    \item 
First, we compare each element with its left neighbor, then if it is necessary we swap them.  \begin{itemize}
    \item This step can be done in parallel, if the swap and comparison can be implemented in $O(poly(\log n))$ number of quantum gates, then it takes $O(poly(\log n))$ time \footnote{Here, since the comparison and swap operations depend on the number of qubits, it may require controlled gates whose decomposition requires number of gates that are polynomial in the system size $O(\log n)$. If this step can be done in $O(1)$, then sorting can be done more efficiently.}.
    \end{itemize} 
\item In the second step we do the same thing for the right neighbors (we group two elements together and compare them.).
 \begin{itemize}
    \item This step also requires $O(poly(\log n))$ number of quantum gates.
\end{itemize} 
\item If we repeat the above steps for O(n) time, we basically get a simple $O(n\log n)$ time sorting algorithm.
\end{itemize}

\section{Conclusion}
In this paper, we describe how to generate suffix structures efficiently as a vector by using quantum circuits for circulant matrices.
We discuss how the generated vector can be used in the string algorithms and sorted if necessary.
As a future direction, we will apply this circuit to the sequence alignment and pattern matching problems.
Since circulant matrices are used in convolutions, it can be also applied to problems in different areas such as convolutional neural network, time series analysis \cite{pollock2002circulant, daskin2022walk}.

\bibliographystyle{IEEEtran}
\bibliography{main}

\begin{thebibliography}{10}
\providecommand{\url}[1]{#1}
\csname url@samestyle\endcsname
\providecommand{\newblock}{\relax}
\providecommand{\bibinfo}[2]{#2}
\providecommand{\BIBentrySTDinterwordspacing}{\spaceskip=0pt\relax}
\providecommand{\BIBentryALTinterwordstretchfactor}{4}
\providecommand{\BIBentryALTinterwordspacing}{\spaceskip=\fontdimen2\font plus
\BIBentryALTinterwordstretchfactor\fontdimen3\font minus
  \fontdimen4\font\relax}
\providecommand{\BIBforeignlanguage}[2]{{%
\expandafter\ifx\csname l@#1\endcsname\relax
\typeout{** WARNING: IEEEtran.bst: No hyphenation pattern has been}%
\typeout{** loaded for the language `#1'. Using the pattern for}%
\typeout{** the default language instead.}%
\else
\language=\csname l@#1\endcsname
\fi
#2}}
\providecommand{\BIBdecl}{\relax}
\BIBdecl

\bibitem{golub2013matrix}
G.~H. Golub and C.~F. Van~Loan, \emph{Matrix computations}.\hskip 1em plus
  0.5em minus 0.4em\relax JHU press, 2013.

\bibitem{gray2006toeplitz}
R.~M. Gray \emph{et~al.}, ``Toeplitz and circulant matrices: A review,''
  \emph{Foundations and Trends{\textregistered} in Communications and
  Information Theory}, vol.~2, no.~3, pp. 155--239, 2006.

\bibitem{karner2003spectral}
H.~Karner, J.~Schneid, and C.~W. Ueberhuber, ``Spectral decomposition of real
  circulant matrices,'' \emph{Linear Algebra and Its Applications}, vol. 367,
  pp. 301--311, 2003.

\bibitem{nielsen2002quantum}
M.~A. Nielsen and I.~Chuang, ``Quantum computation and quantum information,''
  2002.

\bibitem{childs2012hamiltonian}
A.~M. Childs and N.~Wiebe, ``Hamiltonian simulation using linear combinations
  of unitary operations,'' \emph{arXiv preprint arXiv:1202.5822}, 2012.

\bibitem{daskin2012universal}
A.~Daskin, A.~Grama, G.~Kollias, and S.~Kais, ``Universal programmable quantum
  circuit schemes to emulate an operator,'' \emph{The Journal of chemical
  physics}, vol. 137, no.~23, p. 234112, 2012.

\bibitem{low2019hamiltonian}
G.~H. Low and I.~L. Chuang, ``Hamiltonian simulation by qubitization,''
  \emph{Quantum}, vol.~3, p. 163, 2019.

\bibitem{manber1993suffix}
U.~Manber and G.~Myers, ``Suffix arrays: a new method for on-line string
  searches,'' \emph{siam Journal on Computing}, vol.~22, no.~5, pp. 935--948,
  1993.

\bibitem{langmeadburrows}
\BIBentryALTinterwordspacing
B.~Langmead, ``Burrows-wheeler transform and fm index,'' \emph{Johns Hopkins
  University}, accessed in 2022. [Online]. Available:
  \url{https://www.cs.jhu.edu/~langmea/resources/lecture_notes/10_bwt_and_fm_index_v2.pdf}
\BIBentrySTDinterwordspacing

\bibitem{ukkonen1995line}
E.~Ukkonen, ``On-line construction of suffix trees,'' \emph{Algorithmica},
  vol.~14, no.~3, pp. 249--260, 1995.

\bibitem{mielczarek2016review}
M.~Mielczarek and J.~Szyda, ``Review of alignment and snp calling algorithms
  for next-generation sequencing data,'' \emph{Journal of applied genetics},
  vol.~57, no.~1, pp. 71--79, 2016.

\bibitem{burrows1994block}
M.~Burrows and D.~Wheeler, ``A block-sorting lossless data compression
  algorithm,'' in \emph{Digital SRC Research Report}.\hskip 1em plus 0.5em
  minus 0.4em\relax Citeseer, 1994.

\bibitem{hagerup1998sorting}
T.~Hagerup, ``Sorting and searching on the word ram,'' in \emph{Annual
  Symposium on Theoretical Aspects of Computer Science}.\hskip 1em plus 0.5em
  minus 0.4em\relax Springer, 1998, pp. 366--398.

\bibitem{hoyer2002quantum}
P.~H{\o}yer, J.~Neerbek, and Y.~Shi, ``Quantum complexities of ordered
  searching, sorting, and element distinctness,'' \emph{Algorithmica}, vol.~34,
  no.~4, pp. 429--448, 2002.

\bibitem{ambainis2002quantum}
A.~Ambainis, ``Quantum lower bounds by quantum arguments,'' \emph{Journal of
  Computer and System Sciences}, vol.~64, no.~4, pp. 750--767, 2002.

\bibitem{pollock2002circulant}
D.~S.~G. Pollock, ``Circulant matrices and time-series analysis,''
  \emph{International Journal of Mathematical Education in Science and
  Technology}, vol.~33, no.~2, pp. 213--230, 2002.

\bibitem{daskin2022walk}
A.~Daskin, ``A walk through of time series analysis on quantum computers,''
  \emph{arXiv preprint arXiv:2205.00986}, 2022.

\end{thebibliography}
\end{document}